\documentclass[aps,prl,nobibnotes,twocolumn,superscriptaddress,citeautoscript,showkeys]{revtex4-2}

\pdfoutput=1
\usepackage{graphicx}
\usepackage{amsmath, amsfonts}
\usepackage{amssymb}
\usepackage[colorlinks=true,linkcolor=blue,citecolor=blue]{hyperref}
\usepackage[utf8]{inputenc}
\usepackage{float}

\begin{document}

\title[Neural information processing and time-series prediction with only two dynamical memristors]{Neural information processing and time-series prediction with only two dynamical memristors}

\author{D\'{a}niel Moln\'{a}r}
\affiliation{Department of Physics, Institute of Physics, Budapest University of Technology and Economics, M\H{u}egyetem rkp. 3, H-1111 Budapest, Hungary}
\affiliation{ELKH-BME Condensed Matter Research Group, M\H{u}egyetem rkp. 3, H-1111 Budapest, Hungary}

\author{T\'{i}mea N\'{o}ra T\"{o}r\"{o}k}
\affiliation{Department of Physics, Institute of Physics, Budapest University of Technology and Economics, M\H{u}egyetem rkp. 3, H-1111 Budapest, Hungary}
\affiliation{Institute of Technical Physics and Materials Science, Centre for Energy Research, Konkoly-Thege M. \'{u}t 29-33, 1121 Budapest, Hungary}

\author{J\'{a}nos Volk Jr.}
\affiliation{Department of Physics, Institute of Physics, Budapest University of Technology and Economics, M\H{u}egyetem rkp. 3, H-1111 Budapest, Hungary}

\author{Roland K\"{o}vecs}
\affiliation{Department of Physics, Institute of Physics, Budapest University of Technology and Economics, M\H{u}egyetem rkp. 3, H-1111 Budapest, Hungary}

\author{L\'{a}szl\'{o} P\'{o}sa}
\affiliation{Department of Physics, Institute of Physics, Budapest University of Technology and Economics, M\H{u}egyetem rkp. 3, H-1111 Budapest, Hungary}
\affiliation{Institute of Technical Physics and Materials Science, Centre for Energy Research, Konkoly-Thege M. \'{u}t 29-33, 1121 Budapest, Hungary}

\author{P\'{e}ter Bal\'{a}zs}
\affiliation{Department of Physics, Institute of Physics, Budapest University of Technology and Economics, M\H{u}egyetem rkp. 3, H-1111 Budapest, Hungary}

\author{Gy\"{o}rgy Moln\'{a}r}
\affiliation{Institute of Technical Physics and Materials Science, Centre for Energy Research, Konkoly-Thege M. \'{u}t 29-33, 1121 Budapest, Hungary}

\author{Nadia Jimenez Olalla}
\affiliation{Institute of Electromagnetic Fields, ETH Zurich, Gloriastrasse 35, 8092 Zurich, Switzerland}

\author{Zolt\'{a}n Balogh}
\affiliation{Department of Physics, Institute of Physics, Budapest University of Technology and Economics, M\H{u}egyetem rkp. 3, H-1111 Budapest, Hungary}
\affiliation{ELKH-BME Condensed Matter Research Group, M\H{u}egyetem rkp. 3, H-1111 Budapest, Hungary}

\author{J\'{a}nos Volk}
\affiliation{Institute of Technical Physics and Materials Science, Centre for Energy Research, Konkoly-Thege M. \'{u}t 29-33, 1121 Budapest, Hungary}

\author{Juerg Leuthold}
\affiliation{Institute of Electromagnetic Fields, ETH Zurich, Gloriastrasse 35, 8092 Zurich, Switzerland}

\author{Mikl\'{o}s Csontos}
\affiliation{Institute of Electromagnetic Fields, ETH Zurich, Gloriastrasse 35, 8092 Zurich, Switzerland}

\author{Andr\'{a}s Halbritter}\email{halbritter.andras@ttk.bme.hu}
\affiliation{Department of Physics, Institute of Physics, Budapest University of Technology and Economics, M\H{u}egyetem rkp. 3, H-1111 Budapest, Hungary}
\affiliation{ELKH-BME Condensed Matter Research Group, M\H{u}egyetem rkp. 3, H-1111 Budapest, Hungary}

\begin{abstract}
Memristive devices are commonly benchmarked by the multi-level programmability of their resistance states. Neural networks utilizing memristor crossbar arrays as synaptic layers largely rely on this feature. However, the dynamical properties of memristors, such as the adaptive response times arising from the exponential voltage dependence of the resistive switching speed remain largely unexploited. Here, we propose an information processing scheme which fundamentally relies on the latter. We realize simple dynamical memristor circuits capable of complex temporal information processing tasks. We demonstrate an artificial neural circuit with one nonvolatile and one volatile memristor which can detect a neural spike pattern in a very noisy environment, fire a single voltage pulse upon successful detection and reset itself in an entirely autonomous manner. Furthermore, we implement a circuit with only two nonvolatile memristors which can learn the operation of an external dynamical system and perform the corresponding time-series prediction with high accuracy.
\end{abstract}


\date{\today}
\maketitle

\section{Introduction}

Due to their compact size and unsophisticated structure \cite{Pi2019}, low-energy operation \cite{Pickett2012}, high speed \cite{Csontos2023,Witzleben2020,Choi2016a} and CMOS compatibility \cite{Jo2008,Cai2019a,Rao2023}, the applications of memristor synapses in artificial neural networks (ANNs) are booming. Crossbar arrays \cite{Xia2019} of nonvolatile memristors exhibiting multiple resistive states with linear current-voltage [$I(V)$] characteristics \cite{Rao2023,Li2018a} play a key role: The single devices represent the analog synaptic weights, whereas the crossbar array grants the full connectivity between the neighboring neural layers which are typically implemented by standard CMOS circuits \cite{Xia2009,Sheng2019}. Such hybrid CMOS-memristor architectures have been utilized in various neuromorphic computing tasks from data classification \cite{Bayat2018,Li2018b,Wang2018c} to feature extraction \cite{Li2018a,Lin2020} and as field-programmable analog arrays \cite{Li2022}.

Memristor-based synaptic layers efficiently accelerate ANN operation at low energy costs. However, the recursive tuning of the synaptic weights during training still generates extensive software overheads. Alternative approaches exploit the dynamical properties of resistive switching phenomena such as the resistance relaxation of volatile memristors \cite{Du2017,Moon2019} or the exponential dependence of the switching time on the bias voltage \cite{Waser2009,Gubicza2015a} which facilitates short-term/long-term learning and forgetting abilities \cite{Hasegawa2010,Serb2016,Wang2019a}.  As a common figure of merit, ANNs relying on dynamical memristor operation require considerably less network nodes and trained synapses to achieve the same computational efficiency as the more traditional deep neural networks (DNNs) where merely the static properties of the memristors are exploited.
The role of dynamical complexity in the nonlinear, and often adaptive current response of memristors and their small circuits have been revealed recently \cite{Kumar2022}. In a pioneering work, W.~Yi et al. \cite{Yi2018} have faithfully reproduced 23 known neural spiking patterns by utilizing the fourth-order dynamical complexity facilitated by only two coupled VO$_{2}$ memristors. As a next step toward fully memristor-based neuromorphic circuits exhibiting unrivaled footprint, energy efficiency and computational power, the exploration of such dynamical building blocks and the corresponding training algorithms is imperative.

Here, we demonstrate a temporal information processing scheme where a simple circuit consisting of a single nonvolatile memristor implements a complex signal processing device block with a customizable temporal response. As a key component, a negative offset voltage - like artificial forgetting - drives the memristive element towards the initial high-resistance state. In contrast, the analysed input pattern - like learning - pushes it towards low-resistances. A series resistor component introduces a significant voltage division in the sufficiently low-resistance memristive states, which dramatically slows down the response of the memristive element. This simple scheme results in a versatile memristive dynamical system with a very wide tunability of the dynamics and the capability for adjustable short- and long-term memory operations. 

\begin{figure}[t!]
     \includegraphics[width=1\columnwidth]{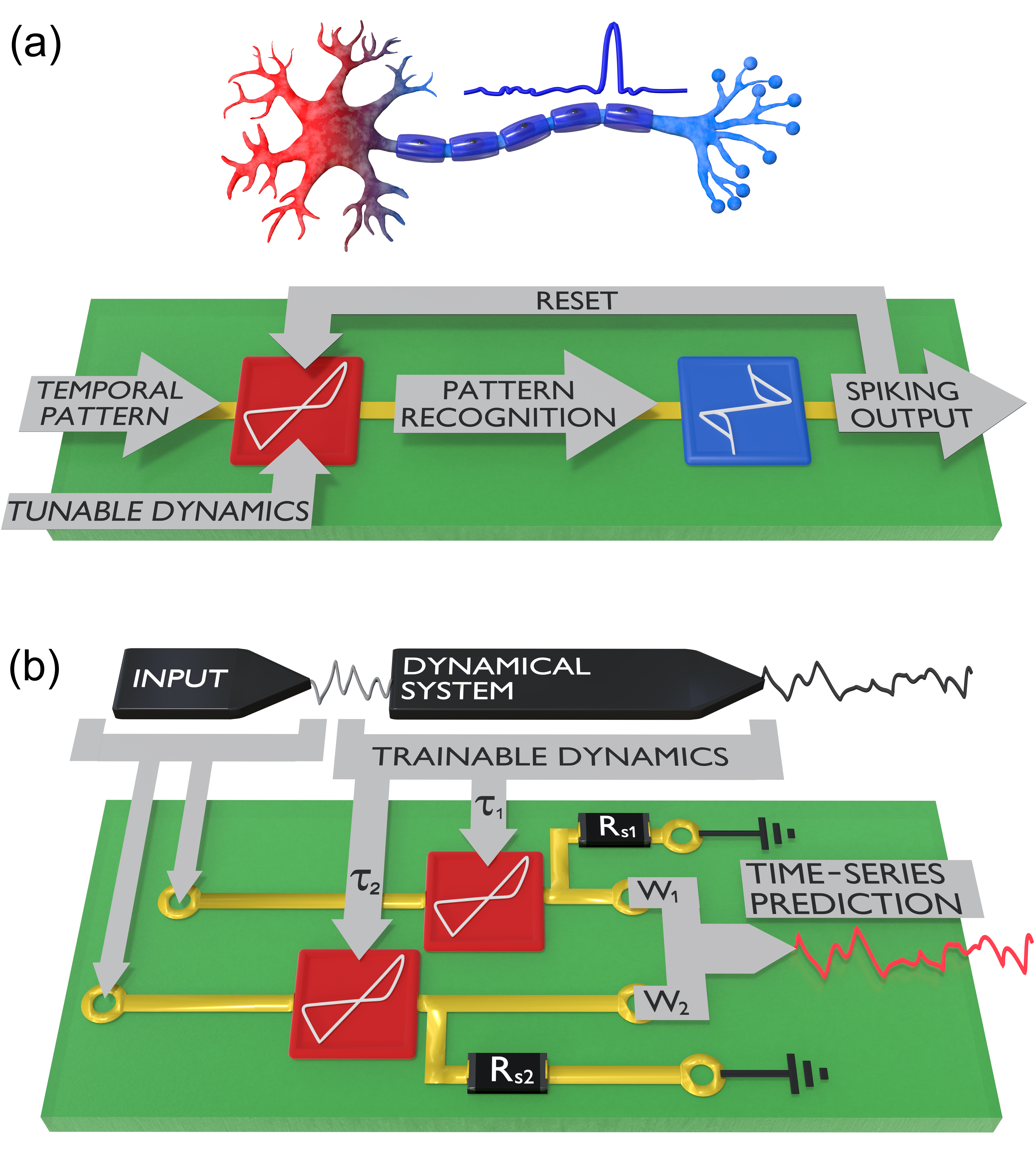}
     \caption{\textbf{Concept of our dynamical memristor circuits.} (a) In analogy to the operation of a biological neuron, a nonvolatile memristor (red) with tunable dynamics analyses an incoming temporal signal. Once the desired signal is recognized, the oscillator circuit of a volatile memristor (blue) starts firing. A feedback to the input resets the non-volatile memristor and thereby stops the firing at the output. (b) Two nonvolatile memristors with tunable dynamics are applied to learn the operation of a dynamical system, and to perform high-accuracy time-series prediction. }
     \label{fig1}
\end{figure}

First, we demonstrate through simulation and experiment that such a small dynamical circuit based on a single nonvolatile Ta$_2$O$_5$ memristor is sufficient to recognize specific temporal patterns such as sub-threshold / super-threshold voltage pulses or neural voltage spikes buried in noise. Next, we extend this detector unit with (i) an oscillator circuit based on a single volatile Mott memristor and (ii) a feedback loop. Thereby we realize a complete artificial neural circuit (Fig.~\ref{fig1}a) which emits a neural voltage spike upon receiving the desired signal. 

Finally, we extend our detection scheme towards circuits with multiple nonvolatile memristors, where each memristor is sensitized to a different dynamical aspect of the input signal stream. This implements a dynamical computing layer that can learn the operation of an external dynamical system and perform time-series prediction\cite{Du2017,Moon2019,LU2024,Hossain2022,Wang2023} as illustrated in Fig.~\ref{fig1}b. Our demonstrator circuit relies on only two Ta$_2$O$_5$ memristors and the linear combination of their output currents. Thanks to the tunable memristor dynamics, it achieves excellent prediction accuracy for the benchmark dynamical system. This result competes with prior pioneering studies, where an array of 90 volatile dynamical memristors was applied to emulate the operation of the same dynamical system with similar accuracy \cite{Du2017}.

Our simple memristive circuits shown in Fig.~\ref{fig1}a,b illustrate the basic operating principle of a scheme that can be generalized to an arbitrary number of input channels with individually tunable dynamics, whose operating range may cover many orders of magnitude, from seconds to nanoseconds.

\section{Results and discussion}


\begin{figure}[t!]
     \includegraphics[width=1\columnwidth]{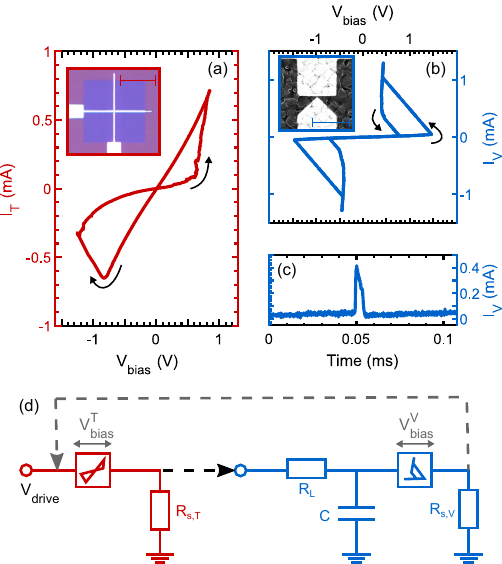}
     \caption{\textbf{The concept of the coupled neural detector \& oscillator circuit.} (a) A representative bipolar current-voltage characteristics of the Ta$_2$O$_5$ memristor measured with a triangular voltage signal of $V_{\rm drive}^{0}$=1.2~V and $f_{\rm drive}$=500~Hz, with $R_{\rm s,T}$=1~k$\Omega$. The inset shows the optical image of the vertically stacked Ta/Ta$_2$O$_5$/Pt crosspoint device. The scale bar corresponds to 50~$\mu$m. (b) A representative unipolar current-voltage characteristics of the VO$_{2}$ memristor measured with a triangle voltage signal of $V_{\rm drive}^{0}$=1.8~V and $f_{\rm drive}$=1~Hz, with $R_{\rm s,V}$=1~k$\Omega$. The inset shows the electron microscope image of the lateral electrode arrangement on top of a VO$_{2}$ layer. The scale bar is 1~$\mu$m. (c) A single voltage spike output of the VO$_{2}$ oscillator circuit shown in (d) in blue. (d) Schematics of the artificial neural circuit. The red icon represents the nonvolatile Ta$_{2}$O$_{5}$ memristor The blue icon stands for the volatile VO$_{2}$ memristor. The module highlighted in red / blue part is responsible for the detector / spiking output functionalities. The lower and upper dashed arrows symbolize the coupling between the two circuit parts and the feedback of the output to the input, respectively. The values of the passive components are given 
     in Supplementary Discussion section S3.}
     \label{fig2}
\end{figure}

\subparagraph{The architecture of the information processing circuits.}

In both of our temporal signal processing schemes (Figs.~\ref{fig1}a,b) we utilize crosspoint Ta/Ta$_2$O$_5$/Pt nonvolatile memristors symbolized by the red icons in Figs.~\ref{fig1}(a,b), and illustrated in the inset of Fig.~\ref{fig2}(a).
A representative current-voltage [$I(V)$] characteristics is shown in Fig.~\ref{fig2}(a). It was measured in the setup highlighted in red in Fig.~\ref{fig2}(d). The $V_{\rm bias}^{\rm T}$ bias voltage corresponding to the voltage drop on the memristor is calculated according to $V_{\rm bias}^{\rm T}=V_{\rm drive}-I_{\rm T}\cdot R_{\rm s,T}$, where $R_{\rm s,T}$ is a series resistor and $V_{\rm drive}$ is the triangular output of the voltage source. Positive voltage refers to a higher potential on the Ta top electrode with respect to the Pt bottom electrode. The hysteretic $I(V)$ trace corresponds to filamentary type resistive switching \cite{Csontos2023,Miao2011,Lee2011a} between metallic states in the analog tunable 1~k$\Omega$\,--\,10~k$\Omega$ regime \cite{Rao2023,Li2018a,Torok2023}.

The neural circuit in Fig.~\ref{fig1}a also utilizes a VO$_2$-based volatile memristor denoted by the blue icon in Fig.~\ref{fig1}(a) and displayed in the inset of Fig.~\ref{fig2}(b).  
Figure~\ref{fig2}(b) exemplifies a typical $I(V)$ curve of the Au/VO$_{2}$/Au memristor, measured in series with the $R_{\rm s,V}$ resistor. This corresponds to unipolar, volatile resistive switching due to a temperature and electric field controlled Mott transition \cite{Valle2019,Posa2023}. In our work, the asymmetric planar layout and the only 30~nm wide lateral gap of the Au electrodes grant a very small and, thus, homogeneous active volume. A detailed experimental analysis and finite element simulations of the device performance and resistive switching mechanism can be found in Ref.~\cite{Posa2023}. In the experiments discussed later, the circuit environment of the Au/VO$_{2}$/Au memristor, highlighted in blue in Fig.~\ref{fig2}(d), is further extended with the capacitor $C$ and the additional resistor $R_{\rm L}$. These, together with the $R_{\rm s,V}$ resistor form a conventional VO$_{2}$ oscillation circuit \cite{Yi2018}. Figure~\ref{fig2}(c) illustrates a single output voltage spike, i.e.\ a single period of the oscillator circuit. Further details on the device preparation of both memristive devices are provided in Supplementary Discussion section S1. 

Figure~\ref{fig2}(d) summarizes the architecture of the neural circuit illustrated in Fig.~\ref{fig1}(a). The lower and upper dashed arrows symbolize the coupling of the input detector and the output firing modules as well as the feedback of the output to the input, respectively. 

The detailed circuit schematics, including the instrumental implementation of the coupling and the feedback is provided in Supplementary Discussion section S3. 
The circuit for time-series prediction (Fig.~\ref{fig1}(b)) relies on the Ta$_2$O$_5$ memristors only, but instead of a single input detector, double Ta$_2$O$_5$ devices with detuned dynamical properties analyze the input stream. 

\subparagraph{The memory time constant of the detector circuit.}

Before presenting the experimentally implemented information processing units, we illustrate the principle of our detection scheme via numerical simulations relying on a minimal model. The key ingredient is the exponential slowdown of the resistive switching as the $V_{\rm bias}^{\rm T}$ bias voltage is linearly decreased on the Ta/Ta$_2$O$_5$/Pt memristor. The exponential voltage dependence of the $\tau$ resistive switching time constant on the bias voltage is commonly known as the `voltage-time dilemma' \cite{Waser2009,Gubicza2015a}. It is a general figure of merit observed in a broad range of resistive switching systems \cite{Chen2017} which describes the dependence of the $I(V)$ characteristics or the pulsed switching response on the speed and voltage amplitude of the driving signal.

In our simulations, we assume $\tau=10^{-\left(\left|V_{\rm bias}^{\rm T}\right|-B\right)/A}$, where the $A$ and $B$ adjustable parameters describe the voltage-dependent dynamical properties, that can be fitted to measured characteristics of our Ta$_2$O$_5$ devices (see Supplementary Discussion section S2).
When a time-dependent $V_{\rm drive}(t)$ driving voltage is applied to the memristor and the $R_{\rm s,T}$ serial resistor, the voltage drop on the memristor is calculated as $V_{\rm bias}^{\rm T}=V_{\rm drive}-I_{\rm T}\cdot R_{\rm s,T}$. The dynamics of the system is determined by a minimal model according to the ${\rm d}x/{\rm d}t=-x/\tau$ and ${\rm d}x/{\rm d}t=(1-x)/\tau$ differential equations for positive/negative bias voltages, respectively. The $0\le x\le 1$ state variable is related to the memristor resistance according to $R_{\rm T}(t)=V_{\rm bias}^{\rm T}/I_{\rm T}=R_{\rm LRS}+\left(R_{\rm HRS}-R_{\rm LRS}\right)\cdot x(t)$, where $R_{\rm HRS}$ ($R_{\rm LRS}$) denotes the highest (lowest) resistance state enabled in our model.

In a highly simplified physical picture, the variation of $x$ can be associated with the voltage-controlled manipulation of the oxygen ion concentration in the filamentary region of the Ta/Ta$_2$O$_5$/Pt memristor \cite{Kumar2022}. In the initial HRS state the poorly conducting filamentary region is saturated with oxygen ions ($x=1$). Due to the applied positive voltage, the mobile oxygen ions are removed from the active region at a rate which is proportional both to the voltage-controlled time constant $\tau$ and the $\sim x$ amount of oxygen ions in the filamentary region. Such a process results in the decrease of the filament resistance and is self-terminated when all the mobile oxygen ions are removed ($x=0$ and $R_{\rm T}=R_{\rm LRS}$). In practice, however, the set process is rather limited by the series resistance, i.e., when $R_{\rm T}(t)$ becomes smaller than $R_{\rm s,T}$ the bias voltage on the memristor decreases, meanwhile the voltage-dependent time constant exponentially increases, and the set transition terminates. At opposite voltage polarity, the oxygen ions are actuated from the surrounding oxide matrix towards the filamentary region, yielding an increase of the resistance. In this case, the rate is again proportional to $\tau$, as well as the $\sim(1-x)$ amount of the available mobile oxygen ions in the active filamentary volume.

\subparagraph{Simulated temporal signal detection.}

\begin{figure}[h!]
     \includegraphics[width=0.98\columnwidth]{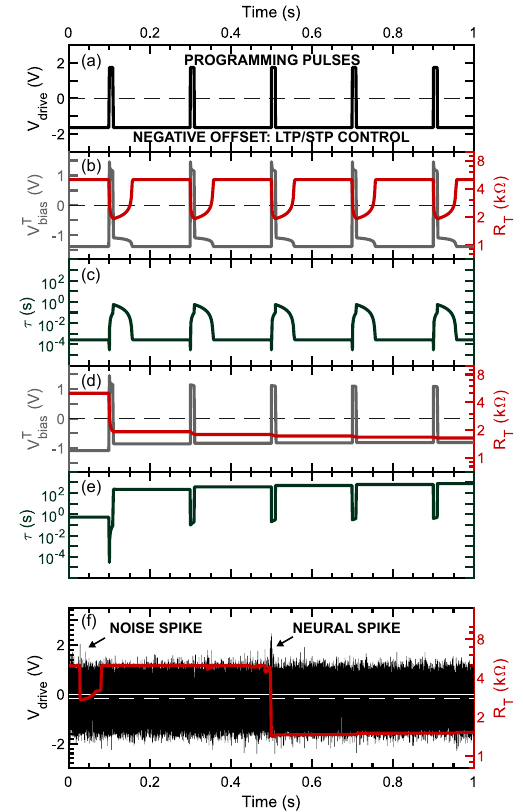}
     \caption{\textbf{Simulation of the voltage-tunable short-term/long-term memory induced in the Ta$_{2}$O$_{5}$ memristor.} (a) The $V_{\rm drive}$ driving voltage signal consists of positive voltage pulses of $V_{\rm pulse}^{0}$=1.75~V peak level, $T_{\rm pulse}$=10~ms duration and 5~Hz repetition rate. Additionally, a constant negative voltage offset of $V_{\rm offset}$ is applied. (b) The red trace shows a short-term memory effect in the time dependence of the $R_{\rm T}$ memristor resistance when a larger magnitude negative voltage offset of $V_{\rm offset}$=-1.65~V is applied. The gray trace shows the time dependence of the $V_{\rm bias}^{\rm T}=V_{\rm drive}-I_{\rm T}\cdot R_{\rm s,T}$ voltage acting on the memristor. (c) The calculated time constant of the resistance change corresponding to the short-term memory effect displayed in (b). (d) The red trace shows a long-term memory effect in $R_{\rm T}$ when a lower magnitude, $V_{\rm offset}$=-1.3~V negative voltage offset is applied while all other parameters are identical to the settings in (b). The gray trace shows the time dependence of the $V_{\rm bias}^{\rm T}$. (e) The calculated time constant of the resistance change corresponding to the long-term memory effect displayed in (d). (f) Simulated resistance response of the Ta$_{2}$O$_{5}$ memristor (red) to an input voltage signal (black) consisting of white noise with 0.45~V rms / 3.6~V pp, a Gaussian voltage spike of $V_{\rm neural}^{0}$= 1.4~V amplitude and 10~ms width as well as a constant negative voltage offset of $V_{\rm offset}$=-0.25~V (white dashed line). $R_{\rm s,T}$=1~k$\Omega$ in all simulations.}
     \label{fig3}
\end{figure}

We apply the above dynamical equations for a single Ta$_2$O$_5$ memristor and a series resistor (red part in Fig.~\ref{fig2}(d)). 
The input pattern scheme illustrated in Fig.~\ref{fig3}(a) consists of positive programming voltage pulses which drive the memristor towards its LRS. In this simple example we use pulses of the same magnitude, but later we show an application where the input information is encoded in the pulse amplitudes. Meanwhile, a constant negative offset voltage is added to the $V_\mathrm{drive}$ input stream, which intends to return the memristor to the initial HRS, and thereby facilitates the forgetting of the programmed information.

The red line in Fig.~\ref{fig3}(b) illustrates a situation, where the first programming pulse decreases the $R_{\rm T}$ memristor resistance to $\approx 2\,\mathrm{k}\Omega$ from its initial $\approx 5\,\mathrm{k}\Omega$ value, but afterwards a sufficient negative offset voltage drives the memristor back to its initial HRS well before the arrival of the next programming pulse. The variation of the $V_\mathrm{bias}^\mathrm{T}$ memristor voltage and the $\tau$ memory time constant during this process are respectively demonstrated by the gray curve in Fig.~\ref{fig3}(b) and Fig.~\ref{fig3}(c).

A fundamentally different behavior is observed when the negative offset voltage is moderately decreased. The first voltage pulse again drives the memristor to an LRS with $\approx 2\,\mathrm{k}\Omega$ (red curve in Fig.~\ref{fig3}(d)). However, due to the smaller negative offset voltage, the memory time constant is too long to reset the memristor before the arrival of the next programming pulse. Note, that the $V_{\rm bias}^{\rm T}=V_{\rm offset}\cdot R_{\rm LRS}/\left( R_{\rm LRS}+R_{\rm s,T}\right)$ voltage division further decreases the portion of the negative offset voltage acting on the memristor (gray curve in Fig.~\ref{fig3}(d)), contributing to the increase of the memory time constant (Fig.~\ref{fig3}(e)) to the range of $100\,s$. The subsequent programming pulses drive the memristor resistance to even smaller values, further increasing the memory time constant. As a result, the memristor gets stuck in the LRS throughout the rest of the input stream.

In this simple example the negative offset voltage introduces a volatile short-term memory behavior in Fig.~\ref{fig3}(b), but this can be turned to a practically non-volatile response to the programming pulses and a related long-term memory operation by simply decreasing the offset voltage (Fig.~\ref{fig3}(d)). We utilize this concept to analyze  more complex input streams by setting the negative offset voltage such that the desired temporal pattern drives the memristor to a long-term memory operation, whereas irrelevant patterns are rapidly forgotten. 

The latter concept is simulated in Fig.~\ref{fig3}f, where the driving signal is a neural spike buried in large noise. This driving signal also includes a negative offset: the white dotted line shows the mean value of the noise, whereas the zero voltage is highlighted by the white solid line. This $V_{\rm offset}$ was optimized such that the resistance of the memristor exhibits only a short-term excursion into a moderately low LRS upon receiving occasional, high-amplitude noise spikes embedded in the white noise input stream. However, when a Gaussian `neural spike' with a comparable amplitude but longer duration arrives, a transition into a virtually permanent lower resistance state is triggered (see the red line). This demonstrates that the dynamics of the memristive signal detection circuit is sensitive to the `strength' of the input signal both in the voltage and time domain: A transition from short-term to long-term memory operation can be induced not only by increasing the voltage amplitude at fixed length but also by increasing the length of the programming signal at a certain voltage amplitude. Thereby the memristor circuit is capable of reliable signal recognition also in situations where traditional voltage threshold-based trigger circuits are challenged by potential noise spikes of short duration but high amplitude.

The optimization of the memory time constant through $V_{\rm offset}$ is similar to a supervised training: one can apply several input streams with and without neural spikes, and modify $V_{\rm offset}$ such that the detection accuracy of the neural spikes is maximized. A similar scheme is applicable to even more complex temporal information processing tasks, like time series prediction (Fig.~\ref{fig1}b): the negative offset values and the related memory time constants of the input channels are optimized for a training dataset in order to achieve maximized prediction accuracy, as will be discussed in Section~2.6.

\subparagraph{Pulse sequence and neural spike detection experiments.}

\begin{figure*}[t!]
     \includegraphics[width=1\textwidth]{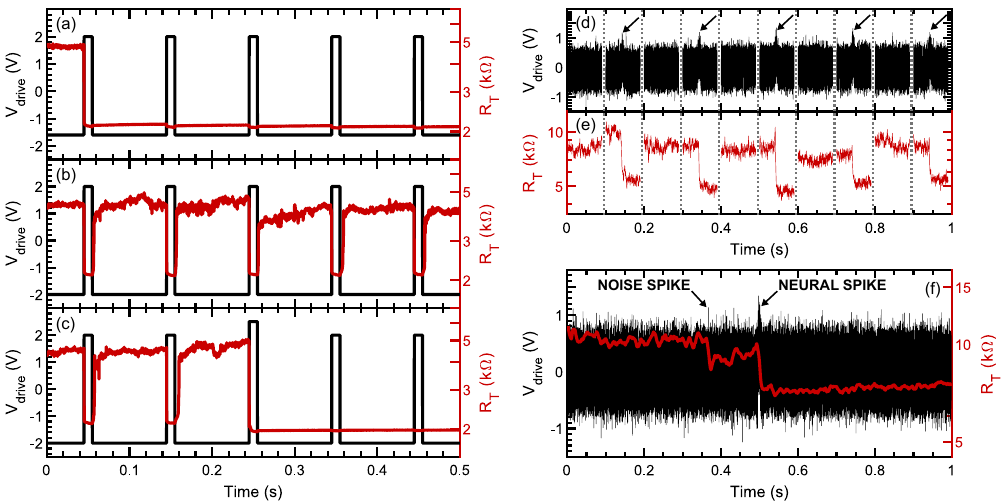}
     \caption{\textbf{Experimental demonstration of the voltage-tunable short-term/long-term memory and neural voltage spike detection using a Ta$_{2}$O$_{5}$ memristor.} (a) Long-term memory response of the $R_{\rm T}$ memristor resistance (red) to an equidistant voltage pulse train of uniform $V_{\rm pulse}^{0}$=2~V peak level, $T_{\rm pulse}$=10~ms duration and 10~Hz repetition rate. The pulses are superimposed at a moderate negative voltage offset of $V_{\rm offset}$=-1.6~V (black). $R_{\rm s,T}$=1~k$\Omega$. (b) Short-term memory response of $R_{\rm T}$ when the magnitude of the negative offset is increased to $V_{\rm offset}$=-2~V whereas the peak value of the input voltage pulses remains $V_{\rm pulse}^{0}$=2~V. (c) Same setting as in (b) with one $V_{\rm drive}$ voltage pulse applied with a higher peak value of $V_{\rm pulse}^{0}$=2.5~V, triggering a long-term response of the Ta$_{2}$O$_{5}$ memristor. (d) 100~ms long sequences of an input voltage signal consisting of white noise with 0.2~V rms / 1.6~V pp and a constant negative voltage offset of $V_{\rm offset}$=-0.042~V. Every second sequence contains a Gaussian voltage spike of $V_{\rm neural}^{0}$=0.4~V and 10~ms width, as marked by the black arrows. (e) The resistance response of the Ta$_{2}$O$_{5}$ memristor to the $V_{\rm drive}$ sequences shown in (d). (f) A 1~s long input voltage sequence of noise, negative voltage offset and a single Gaussian voltage pulse, labeled as `neural spike' (black), utilizing the same parameters as in (d). A larger noise spike is also labeled. The resistance response exhibits a short-term / long-term memory effect induced by the noise spike / neural spike. In the presence of the input noise $R_{\rm T}$ was calculated from the 1~ms long moving averages of the Ta$_{2}$O$_{5}$ memristor's current and bias voltage.}
     \label{fig4}
\end{figure*}

The simulation results of the signal detection scheme were experimentally verified by using a single Ta$_2$O$_5$ memristor and a $R_{\rm s,T}$=1~k$\Omega$ series resistor, as shown in Fig.~\ref{fig4}. Long- and short-term memory operation due to uniform, positive voltage pulses over a constant negative background are demonstrated in Fig.~\ref{fig4}(a) and (b), respectively. Here the peak values of the programming pulses as well as the initial conditions of the circuit are identical, except for the negative voltage offset which was changed from $V_{\rm offset}$=-1.6~V (long-term memory) to $V_{\rm offset}$=-2~V (short-term memory). The selectivity of the short-term / long-term response of the circuit to the programming voltage amplitude was tested by exposing it to the input pattern consisting of uniform programming pulses except for one which was 0.5~V higher in its peak value, as shown in Fig.~\ref{fig4}(c). Here, the negative offset is optimized to erase the long-term effect of sub-threshold pulses, but the third, higher pulse drives the system to a long-term memory state. This observation implies that the circuit can function as a detector module which recognizes super-threshold input signals by the transition into an enduring LRS of the Ta$_2$O$_5$ memristor whereas sub-threshold signals only induce short-term, fully recovering resistance changes.

Neural spike detection in a noisy environment, as proposed by the simulations shown in Fig.~\ref{fig3}(f), has been tested experimentally by applying independent time series of identical white noise characteristics, a negative voltage offset and occasional neural spikes, i.e., Gaussian voltage pulses to the detector circuit. Example time traces of such input streams are illustrated in Fig.~\ref{fig4}(d), where every second sequence contains a Gaussian spike, as labeled by the arrows. The measured resistance response of the memristor, shown in Fig.~\ref{fig4}(e) verifies that the neural spikes always trigger the transition into an enduring LRS whereas random noise only results in quickly fading, smaller resistance changes. This contrast is better exemplified in Fig.~\ref{fig4}(f), where a selected single time trace contains both a higher-amplitude, short noise spike and a comparable amplitude, but longer lasting neural spike. The former induces a smaller magnitude, quickly fading resistance drop while the latter facilitates the transition to an enduring LRS.

For a statistical analysis 50-50 noisy time-series including/lacking a neural spike were experimentally evaluated by the detector circuit at nominally identical conditions, assuming an optimized $V_{\rm offset}$. As successful detection events only those time series are considered which (i) contain a neural spike and only this spike is detected at its true position or (ii) there is no neural spike in the stream and none is detected, accordingly. The detection accuracy, defined as the ratio of the successfully classified streams was found to be 99\% at 0.4~V pulse amplitude and 0.2~V rms noise {(see Supplementary Discussion section S4 for similar analysis at different signal-to-noise ratios)}. According to the rms to peak-to-peak (pp) conversion of white noise, this rms value yields $<$1.6V peak-to-peak fluctuations in 98\% of the time. In comparison, an alternative approach based solely on the threshold switching property of metal-oxide memristors yielded in $\approx$60\% detection accuracy at $\approx$1.7~V spike amplitude and 0.21~V rms / 1.1~V pp noise levels \cite{Gupta2016,Gupta2017,Gupta2019}.

\subparagraph{Neural transceiver operation.}

\begin{figure*}[t!]
     \includegraphics[width=1\textwidth]{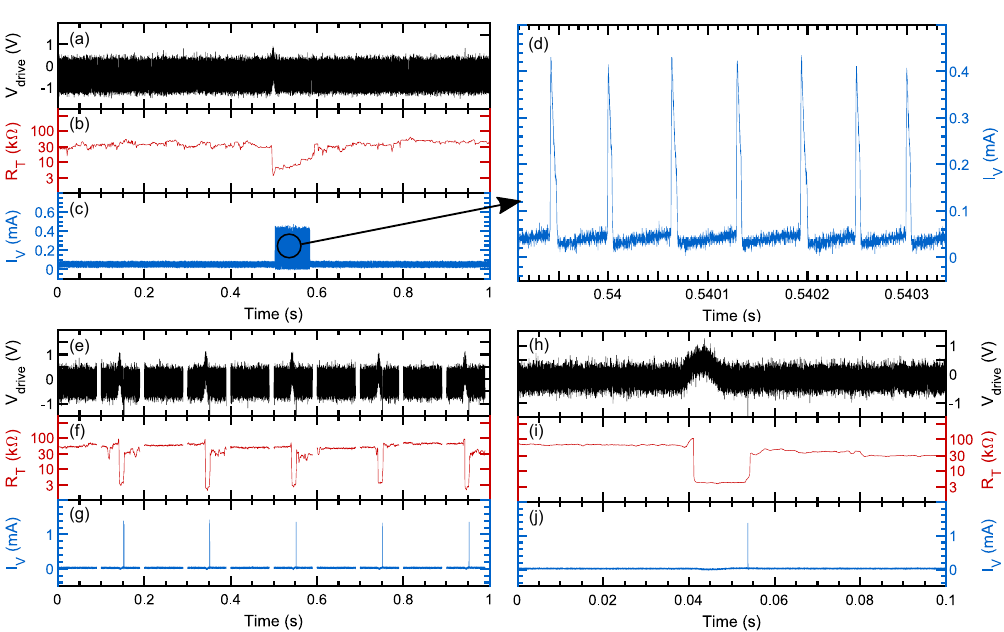}
     \caption{\textbf{The operation of the coupled neural detector \& oscillator circuit.} (a-d) Establishing the coupling symbolized by the lower dashed arrow in Fig.~\ref{fig2}(d). (a) $V_{\rm drive}$ input signal sequence consisting of white noise with 0.24~V rms / 1.92~V pp, a negative voltage offset of $V_{\rm offset}$=-0.46~V and a single Gaussian voltage spike of 0.47~V amplitude and 10~ms width. (b) The corresponding $R_{\rm T}$ resistance response of the Ta$_{2}$O$_{5}$ memristor. $R_{\rm s,T}$=1~k$\Omega$. (c) When the $R_{\rm T}$ resistance drops upon the spike detection, the voltage change on the $R_{\rm s,T}$ series resistor shifts the operation point of the VO$_{2}$ module, triggering periodic $I_{\rm V}$ current oscillations at its output. The oscillations persist for a period determined by the memory time constant of the detector module. $R_{\rm L}$=50~k$\Omega$, $R_{\rm s,V}$=1~k$\Omega$, $C$=4.4~nF. (d) Magnified view of the $I_{\rm V}$ current oscillations shown in (c). (e-j) Adding the global feedback symbolized by the upper dashed arrow in Fig.~\ref{fig2}(d). (e) 100~ms long sequences of an input voltage signal consisting of white noise with 0.22~V rms / 1.76~V pp and a constant negative voltage offset of $V_{\rm offset}$=-0.15~V. Every second sequence contains a Gaussian voltage spike of 0.5~V amplitude and 10~ms width. (f) The $R_{\rm T}$ resistance of the Ta$_{2}$O$_{5}$ memristor undergoes a set switching which triggers the oscillatory behavior of the firing circuit. However, the coupling of the first oscillation at the output to the input of the detector circuit restores the HRS of the Ta$_{2}$O$_{5}$ memristor and, thus, terminates the current oscillation at the output after its first period. (h-j) Magnified views of $V_{\rm drive}$, $R_{\rm T}$ and $I_{\rm V}$, respectively, during a single input sequence utilizing the same parameters as in (e). Note, that the $I_{\rm V}$ current spike is delayed with respect to the set transition of $R_{\rm T}$ due to the low-pass filter established in the coupling (see Supplementary Discussion section S3).} 
     \label{fig5}
\end{figure*}

In order to add the transceiver functionality and automated reset to our Ta$_{2}$O$_{5}$-based neural spike detection circuit, we extend it by a VO$_{2}$-based neural spike firing module and a global feedback loop, according to the blue shaded layout parts and the gray dashed arrow in Fig.~\ref{fig2}(d), respectively. The detailed circuit implementation is outlined in Supplementary Discussion section S3. 
First, we discuss the behavior of the zero-feedback circuit which actuates a firing output upon spike detection. Next, the global feedback is activated, completing the circuit operation with output optimization and automated reset.

Figures~\ref{fig5}(a-d) exemplify an experiment utilizing the input stream displayed in Fig.~\ref{fig5}(a), consisting of white noise, a single neural spike and a constant negative voltage offset. The strength of the neural spike and the negative voltage offset are adjusted such that the memory time constant of the detector circuit is in the order of $\approx$100~ms, as shown in Fig.~\ref{fig5}(b). While the Ta$_{2}$O$_{5}$ memristor resides in its LRS, the voltage drop on the $R_{\rm s,T}$ series resistor, arising from the constant negative $V_{\rm offset}$, is increased in magnitude due to the voltage divider effect of $R_{\rm T}$ and $R_{\rm s,T}$. We utilize this voltage drop as the input signal of the VO$_{2}$ oscillator module. The coupling between the detector and oscillator modules is realized through a differential amplifier which sets the appropriate input level for the VO$_{2}$ circuit and also acts as a low-pass filter. This filtering implements an integration time constant that averages out the rapidly changing signal components, and practically measures the effect of the negative offset voltage on the $R_{\rm s,T}$ series resistor. The latter exhibits an order of magnitude increase upon the set transition of the Ta$_{2}$O$_{5}$ memristor. While the Ta$_{2}$O$_{5}$ memristor is in its HRS, the oscillation module is in its steady state, maintaining a constant, low-level $I_{\rm V}$ output current corresponding to the HRS of the VO$_{2}$ memristor. However, when the Ta$_{2}$O$_{5}$ circuit detects a neural spike, the increased voltage input of the VO$_{2}$ circuit results in the firing of periodic current pulses, as shown in Fig.~\ref{fig5}(c). This so-called tonic spiking pattern, magnified in Fig.~\ref{fig5}(d) is explained in terms of the periodic charging and discharging of the parallel capacitor $C$ due to the current supplied through $R_{\rm L}$ and the voltage-induced current instability of the VO$_{2}$ memristor, respectively \cite{Yi2018}. The conditions of the periodic firing are maintained as long as the fading LRS of the Ta$_{2}$O$_{5}$ resides below a threshold, determined by $V_{\rm offset}$, $R_{\rm s,T}$, the voltage gain of the coupling amplifier and the $I(V)$ characteristics of the VO$_{2}$ memristor. The rise and fall times as well as the periodicity of the individual $I_{\rm V}$ output current pulses are tunable via the $R_{\rm L}$, $C$ and $R_{\rm s,V}$ component values.

Finally, the neuro-transceiver circuit is completed by introducing the global feedback from its spiking output to the detector input. This is realized via a second differential amplifier such that the positive valued output spikes of the VO$_{2}$ module appear with a negative polarity in the $V_{\rm drive}$ detector input stream. At the same time, they are amplified to a level where a single negative spike is sufficient to trigger the reset transition of the Ta$_2$O$_5$ memristor in a short time compared to the repetition rate of the output spikes. As a result, the oscillator module is brought back to its steady state and the neuro-transceiver circuit is ready to receive, detect and flag the next neural input spike (see Supplementary Discussion section S3).

The reliable operation of the above scheme is experimentally demonstrated in Figs.~\ref{fig5}(e-g) where the circuit is exposed to 10 independent input streams consisting of white noise and a constant negative offset. The latter now incorporates a portion of the default output through the global feedback loop. Every second input sequence also contains a Gaussian neural spike, as shown by the black traces in Fig.~\ref{fig5}(e). The resistance response of the Ta$_{2}$O$_{5}$ memristor is displayed in Fig.~\ref{fig5}(f) in red. During the input sequences lacking a neural signal $R_{\rm T}$ stays in the HRS and the corresponding $I_{\rm V}$ current output of the VO$_{2}$ module, shown in Fig.~\ref{fig5}(e), also remains at a constant low level. However, whenever a neural spike arrives at the input, $R_{\rm T}$ drops to the LRS, an output spike is triggered and the HRS of $R_{\rm T}$ is restored. 
The set transition of $R_{\rm T}$ (red) occurs instantaneously during the $\approx$5~ms rise time of the neural spike (black). The output spike (blue) is fired 13~ms later. This delay corresponds to the elongated rise time of the detector module's output at the firing module's input due to the low-pass filter implemented in their coupling. The output spike simultaneously appears with a negative polarity at the input (black) through the feedback amplifier and facilitates the reset of the Ta$_{2}$O$_{5}$ memristor and, thus, the whole circuit within 1~ms.

Note, that the time constant of the low-pass filter exceeds not only the one of the random fluctuations but also the duration of the neural spikes. This grants that neither the noise nor the neural spikes can directly trigger the firing of the VO$_{2}$ circuit. The latter is only facilitated by the set transition of the Ta$_{2}$O$_{5}$ memristor, whose latching operation improves the reliability of the neuron circuit. Moreover, the filtering allows a longer `integration' of the input stream by the detector circuit, resulting in higher detection accuracy. The cut-off frequency of the coupling and, thus, the integration time of the detector circuit are instrumentally optimized to suit the specific input pattern.

The statistical evaluation of the detection accuracy of the complete neuro-transceiver was carried out as described for the detector circuit in Section~2.3. In comparison to the latter, the experimentally verified detection accuracy was slightly decreased from 99\% to 98\% at comparable signal-to-noise ratio characterized by 0.5~V pulse amplitude and 0.22~V rms / 1.76~V pp noise levels. The additional detection failure was identified to arise from the partial set transition of the Ta$_{2}$O$_{5}$ memristor at a true pulse detection event which was insufficient to access the oscillatory operation regime of the VO$_{2}$ memristor.

\subparagraph{Time-series prediction with two non-volatile Ta$_2$O$_5$ memristors.}



Finally, we move forward from temporal signal recognition and utilize our dynamical memristor circuits for time-series prediction. {Memristive reservoirs were already utilized for various time-series prediction benchmark problems \cite{Du2017,jang2024,Zhong2021,Choi2024,Moon2019,LU2024,Hossain2022,Wang2023,TANAKA2019}, like Logistic and Hénon maps, Mackey-Glass and Nonlinear Autoregressive Moving-average (NARMA) time series, and time-dependent Lorenz attractors. Common to these approaches is the use of a large number of memristors in the reservoir, or alternatively external processing or storage functionalities are applied to reduce the complexity of the reservoir itself. These include customized temporal rescaling of the inputs~\cite{Du2017}, preprocessing masks~\cite{jang2024,Zhong2021,LU2024,Hossain2022} or virtual nodes~\cite{jang2024,Zhong2021,Moon2019,LU2024} for the partial storage of the memristors' state-evolution in conventional memories. Depending on the actual arrangement, these memristive time-series prediction architectures include several tens or several hundreds of memristive elements in the reservoir. Here, we provide a proof-of-principle demonstration that a very simple architecture with only two memristive channels (Fig.~\ref{fig1}(b)) can already solve a relatively complex time-series prediction task, once the operation dynamics of the memristive channels can be tailored for the targeted problem.}

\begin{figure}[t!]
             \includegraphics[width=\columnwidth]{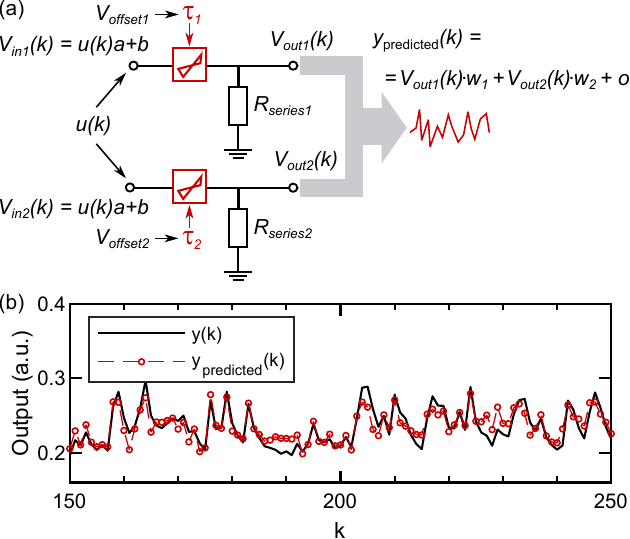}
     \caption{{\textbf{Time-series prediction of a dynamical system using two Ta$_2$O$_5$ memristors.} (a) The scheme of the analyis. After proper scaling, the $u(k)$ mathematical input pattern is applied to the inputs of the two memristive channels in the form of positive voltage pulses. Each channel includes a non-volatile Ta$_2$O$_5$ memristor (red symbols) and a series resistor. The $\tau_1$ and $\tau_2$ memory time-constants of the channels are adjustable through the applied negative offset voltages. The response of the channels is read out through the $V_\mathrm{out1}$ and $V_\mathrm{out2}$ output voltages measured along the readout voltage segments of the driving stream (see the full driving scheme in Fig.~S5 of the Supplementary Discussion section S5). Finally, the predicted output is calculated as a linear combination of the output voltages. (b) The $y_\mathrm{predicted}(k)$ predicted output stream of the memristive circuit (red circles) in comparison to the true $y(k)$ output of the mathematical dynamical system (black line). Following the visualization format in Ref.~\citenum{Du2017}, here only a shorter segment of the validation data-stream is plotted for better visibility, and the the full validation data is available in the Supplementary Discussion section S5 (Supplementary Fig.~S6).}
}
     \label{fig6}
\end{figure}

For this purpose, we investigate the same {NARMA2} benchmark nonlinear dynamical system which was used in the pioneering work of C.~Du and coworkers \cite{Du2017}, where a memristive reservoir computing layer of volatile WO$_x$ devices performed the prediction task. This dynamical system relies on the $y(k)=0.4y(k-1)+0.4y(k-1)y(k-2)+0.6u^3(k)+0.1$ equation \cite{Atiya2000}, where $y(k)$ is the $k$th output state of the system in response to the series of $u(k)$ uniformly distributed random inputs. Note, that this equation includes nonlinear functions and the actual $y(k)$ output also depends on the previous and previous-to-previous output states ($y(k-1)$ and $y(k-2)$). An example output stream of this dynamical equation is exemplified by the black solid line in Fig.~\ref{fig6}b.

The memristive dynamical network under test has to learn the operation of this dynamical system such, that it converts the $u(k)$ inputs to a $y_\mathrm{predicted}(k)$ predicted output stream, which mimics the true $y(k)$ output. First a training is performed, where the parameters of the memristor circuit are tuned such, that the $\mathrm{NMSE}=\sum_k\left(y_\mathrm{predicted}(k)-y(k)\right)^2/\sum_k\left(y^2(k)\right)$ normalized mean squared error is minimized for a training dataset. Afterwards, the optimized parameters of the memristive network are set and the system autonomously predicts the output of the dynamical system for any input pattern. This operation is characterized by the NMSE$_\mathrm{validation}$ evaluated on a validation dataset.

In Ref.~\citenum{Du2017} a reservoir computing layer with 90 volatile WO$_x$ memristors was utilized. These memristive units exhibited a non-tunable, $\approx 50\,$ms memory time-constant. To scale the $u(k)$ input values to the optimal operation range of the memristive units a $V_{\rm drive}(k)=u(k)\cdot a+b$ linear transformation was applied, which used the same $a$ amplification and $b$ offset for all the 90 memristive inputs. Furthermore, the input stream was fed to the memristive layer using 10 different temporal scaling factors. The latter temporal transformation artificially introduced a variation of the effective time constants. The $y_\mathrm{predicted}(k)$ predicted output of the memristive network was calculated as the linear combination of the individual memristive output states (memristor resistances) using 90 trainable weights and an additional offset value. At optimized parameters a normalized mean squared error of NMSE$_\mathrm{training}=3.61\cdot10^{-3}$ was obtained on the training dataset. Finally, the prediction error was tested on an independent validation dataset yielding NMSE$_\mathrm{validation}=3.13\cdot10^{-3}$. 



Our approach (see Fig.~\ref{fig1}(b) and Fig.~\ref{fig6}) relies on the dynamical operation of only two nonvolatile Ta/Ta$_2$O$_5$/Pt memristor units exploiting their tunable memory time constants. The driving scheme was similar to Fig.~\ref{fig3}a with the difference, that non-identical positive voltage pulses represented the input signal ($u(k)$) to both memristors,
whereas a constant negative offset voltage  was applied in-between the programming pulses
to facilitate forgetting. {In practice, the $u(k)\in [0,0.5]$ uniformly distributed input random numbers were linearly scaled to $V_\mathrm{in1}(k)$ and $V_\mathrm{in2}(k)$ driving pulse amplitudes by applying the same $a=1.5\,\mathbf{V}, b=1\,\mathbf{V}$ scaling factors for the two inputs (see Fig.~\ref{fig6}a) such that the input stream is properly adjusted to the operation range of the memristive units. The relaxation dynamics of the latter was intentionally de-tuned: the two forgetting time constants ($\tau_1$ and $\tau_2$) were granted by different negative offset voltages ($V_\mathrm{offset1}$ and $V_\mathrm{offset2}$, see Fig.~\ref{fig6}a). The outputs of the memristive circuits were read out as the $V_\mathrm{out1}(k)$ and $V_\mathrm{out2}(k)$ voltage drops on the $R_\mathrm{series1}=R_\mathrm{series2}=1050\,\mathrm{\Omega}$ series resistors. This can be done by introducing a $200\,$mV read voltage pulse between the programming pulse and the negative offset (see the full driving scheme in Fig.~S5 of the Supplementary Discussion section S5). The $V_\mathrm{out1}(k)$ and $V_\mathrm{out2}(k)$ voltage values measured during the read pulse are fed directly into the readout layer.} 
Finally, the predicted output is calculated as a linear combination of the output voltages using two weights and an offset ($w_1$, $w_2$, $o$, see Fig.~\ref{fig6}a).
The latter step was completed in software. 

{During the experimental demonstration, first a training dataset with $300$ random numbers ($u_\mathrm{training}(k)$) was fed to the system. The individual voltage responses $V_\mathrm{out1}(k)$ and $V_\mathrm{out2}(k)$ were measured at various offset parameters systematically scanning the physically relevant $1.5\,\mathrm{V}<V_\mathrm{offset1},V_\mathrm{offset2}<2.1\,\mathrm{V}$ range with $0.05\,\mathrm{V}$ resolution for both memristive units. At a certain offset value pair the adjustable parameters of the read-out linear combination ($w_1$, $w_2$, $o$) were optimized in software such that the best NMSE is obtained at the given hardware parameter set. Finally, the globally best setting was selected by choosing the hardware parameter set at which the overlall best NMSE$_\mathrm{training}=3.09\cdot10^{-3}$ was obtained.} 

To validate the trained system, the prediction accuracy was evaluated on an independent validation dataset ($u_\mathrm{validation}(k), k=1..300$) with the best performing training parameters, obtaining NMSE$_\mathrm{validation}=3.03\cdot10^{-3}$. This is demonstrated in Fig.~\ref{fig6}b, where the red circles exhibit the $y_\mathrm{predicted}(k)$ predicted data in comparison to the true $y(k)$ response. 
{These NMSE values represent a similar predictive performance to Ref.~\citenum{Du2017} using significantly less physical units and tunable system parameters. Further details, including the training and validation input streams, the training data, the best performing system parameters, and further optimization possibilities are available in the Supplementary Discussion section S5.} 



These results show that a very simple dynamical layer with only two nonvolatile memristors is sufficient to predict a complex dynamical system with high accuracy. {Such a minimal number of hardware components is enabled by the wide tunability of the system dynamics through the independent adjustment of the forgetting time of each unit.}
This way the system parameters can be efficiently customized to hardware-encode the solver of the targeted dynamical problem. This is in contrast to reservoir layers relying on non-tunable, hardware-defined dynamical properties lacking the adaptability to the problem, necessitating software-based solutions such as the off-line temporal re-scaling of the input data. Consequently, our architecture allows real-time hardware operation, provided that the read-out weights are also hardware-implemented with memristive artificial synapses. Adding additional input channels is nonetheless straightforward, if the complexity of the task requires even more individually tunable dynamical channels.

\section{Conclusion}

In conclusion, we have demonstrated an information processing scheme, where a nonvolatile memristor is programmed by positive voltage patterns whereas a negative offset voltage induces forgetting. This driving pattern, together with the use of a series resistor, enables a broad tunability of the resistive switching dynamics, including adjustable transition between short- and long-term memory operation. We have shown that with this approach, a single memristior is sufficient for performing temporal signal recognition tasks such as discriminating between sub-threshold and super-threshold voltage pulses or detecting neural voltage spikes in noisy environment.

If an additional input module (another nonvolatile memristor and another series resistor) is added, the two memristors can be sensitized to different dynamical aspects of the input signal stream and, accordingly, more complex temporal patterns can be analyzed. With this architecture, we have achieved high-accuracy time-series prediction. The adjustable dynamical properties were exploited in the training phase. As a result, the simplistic circuit consisting of only two memristors, two series resistors and a suitably chosen linear combination of the individual output voltages was able to predict the output of a benchmark dynamical system with high accuracy.

We emphasize, that the relaxation time constant towards the HRS is not an entirely hardware-encoded physical property in our scheme. Instead, the forgetting rate can be adjusted by the negative offset voltage. Here, we investigated the voltage dependence of the switching speed over a limited frequency range ($1\,$Hz-$1\,$kHz, see Supplementary Figure~S1c,d). However, our high-frequency study on similarly prepared Ta$_2$O$_5$ memristors revealed set (reset) times down to 10~ps (300~ps) with a clear voltage-dependent switching times at these extreme speeds \cite{Csontos2023}. This can extend our signal analysis scheme by up to 10 orders of magnitude in the time domain, in contrast to those reservoir computing solutions where a fixed or less tunable physical relaxation time constant is utilized. We believe that the straightforward extension of our system with an arbitrary number of independently adjustable input modules offers a versatile platform for a wide range of temporal signal analysis tasks. 


Furthermore, we have shown that our temporal signal analysis scheme can be extended with additional hardware components that post-process the output of the nonvolatile memristor layer. To this end, we coupled the oscillator circuit of a volatile VO$_2$ Mott memristor to the output of the non-volatile Ta$_2$O$_5$ memristor. Thereby we implemented a complete neuro-transceiver unit which detects a neural spike buried in high noise, emits a high-amplitude neural spike upon detection, and resets the input detector layer in a fully autonomous manner.

With these results, we demonstrate that simplistic memristive circuits with very few components can efficiently solve high-complexity temporal signal processing tasks, once the rich dynamical properties of the memristors are flexibly exploited. This possibility offers highly compact and efficient building blocks for those edge computing tasks where energy-hungry software solutions must be avoided.

\section*{Acknowledgements}

This research was supported by the Ministry of Culture and Innovation and the National Research, Development and Innovation Office within the Quantum Information National Laboratory of Hungary (Grant No. 2022-2.1.1-NL-2022-00004), and the NKFI K143169, K143282 and TKP2021-NVA-03 grants. Project no. 963575 has been implemented with the support provided by the Ministry of Culture and Innovation of Hungary from the National Research, Development and Innovation Fund, financed under the KDP-2020 funding scheme. L.P. and Z.B. acknowledge the the support of the Bolyai J\'{a}nos Research Scholarship of the Hungarian Academy of Sciences. J.L., M.C. and N.J.O. acknowledge the financial support of the Werner Siemens Stiftung.

\section*{Author contributions}

The Ta$_{2}$O$_{5}$-based temporal signal recognition experiments were developed and performed by D.M. The experiments with the combined Ta$_{2}$O$_{5}$ and VO$_{2}$ circuit were carried out by D.M., T.N.T. and R.K. The time-series prediction measurements were performed and analyzed by D.M. and J.V.Jr. The simulations were initiated by the work of P.B. and were extended and finalized by D.M. The Ta$_{2}$O$_{5}$ memristors were developed and fabricated by M.C. and N.J.O. in the group of J.L. The VO$_{2}$ thin layers were manufactured and optimized by G.M. and J.V., the VO$_{2}$ devices were developed and manufactured by L.P., and T.N.T. in the group of J.V.. The project was conceived and supervised by A.H. The manuscript was written by M.C, D.M., T.N.T. and A.H. with contributions from Z.B. All authors contributed to the discussion of the results.

\section*{Competing interests}
The authors declare no competing interests.


\bibliography{References.bib}

\end{document}


\title{Neural information processing and time-series prediction with only two dynamical memristors\\ \vspace{0.5cm}
Supplementary Information \vspace{0.2cm}}

\author{Dániel Molnár}
\affiliation{Department of Physics, Institute of Physics, Budapest University of Technology and Economics, M\H{u}egyetem rkp. 3., H-1111 Budapest, Hungary.\looseness=-1}
\affiliation{ELKH-BME Condensed Matter Research Group, M\H{u}egyetem rkp. 3., H-1111 Budapest, Hungary.\looseness=-1}

\author{Tímea Nóra Török}
\affiliation{Department of Physics, Institute of Physics, Budapest University of Technology and Economics, M\H{u}egyetem rkp. 3., H-1111 Budapest, Hungary.\looseness=-1}
\affiliation{Institute of Technical Physics and Materials Science,\unpenalty~Centre for Energy Research, Konkoly-Thege M. \'{u}t 29-33, 1121 Budapest, Hungary.\looseness=-1}

\author{János Volk Jr.}
\affiliation{Department of Physics, Institute of Physics, Budapest University of Technology and Economics, M\H{u}egyetem rkp. 3., H-1111 Budapest, Hungary.\looseness=-1}

\author{Roland Kövecs}
\affiliation{Department of Physics, Institute of Physics, Budapest University of Technology and Economics, M\H{u}egyetem rkp. 3., H-1111 Budapest, Hungary.\looseness=-1}

\author{László Pósa}
\affiliation{Department of Physics, Institute of Physics, Budapest University of Technology and Economics, M\H{u}egyetem rkp. 3., H-1111 Budapest, Hungary.\looseness=-1}
\affiliation{Institute of Technical Physics and Materials Science,\unpenalty~Centre for Energy Research, Konkoly-Thege M. \'{u}t 29-33, 1121 Budapest, Hungary.\looseness=-1}

\author{Péter Balázs}
\affiliation{Department of Physics, Institute of Physics, Budapest University of Technology and Economics, M\H{u}egyetem rkp. 3., H-1111 Budapest, Hungary.\looseness=-1}

\author{György Molnár}
\affiliation{Institute of Technical Physics and Materials Science,\unpenalty~Centre for Energy Research, Konkoly-Thege M. \'{u}t 29-33, 1121 Budapest, Hungary.\looseness=-1}

\author{Nadia Jimenez Olalla}
\affiliation{Institute of Electromagnetic Fields, ETH Zurich, Gloriastrasse 35, 8092 Zurich, Switzerland.\looseness=-1}

\author{Zoltán Balogh}
\affiliation{Department of Physics, Institute of Physics, Budapest University of Technology and Economics, M\H{u}egyetem rkp. 3., H-1111 Budapest, Hungary.\looseness=-1}
\affiliation{ELKH-BME Condensed Matter Research Group, M\H{u}egyetem rkp. 3., H-1111 Budapest, Hungary.\looseness=-1}

\author{János Volk}
\affiliation{Institute of Technical Physics and Materials Science,\unpenalty~Centre for Energy Research, Konkoly-Thege M. \'{u}t 29-33, 1121 Budapest, Hungary.\looseness=-1}

\author{Juerg Leuthold}
\affiliation{Institute of Electromagnetic Fields, ETH Zurich, Gloriastrasse 35, 8092 Zurich, Switzerland.\looseness=-1}

\author{Miklós Csontos}
\affiliation{Institute of Electromagnetic Fields, ETH Zurich, Gloriastrasse 35, 8092 Zurich, Switzerland.\looseness=-1}

\author{András Halbritter}\email{halbritter.andras@ttk.bme.hu}
\affiliation{Department of Physics, Institute of Physics, Budapest University of Technology and Economics, M\H{u}egyetem rkp. 3., H-1111 Budapest, Hungary.\looseness=-1}
\affiliation{ELKH-BME Condensed Matter Research Group, M\H{u}egyetem rkp. 3., H-1111 Budapest, Hungary.\looseness=-1}

\maketitle
\thispagestyle{fancy}
\lhead[]{Supplementary Information}
\rhead[]{}

\newpage

\setcounter{secnumdepth}{1}

\section{Sample fabrication}

The 10~nm thick Ti adhesive layer and the 40~nm thick Pt bottom electrode of the Ta/Ta$_{2}$O$_{5}$/Pt memristor were subsequently deposited on a 280~nm thick SiO$_{2}$ substrate by electron beam evaporation at a base pressure of 10$^{-7}$~mbar at a rate of 0.1~nm/s. The 5~nm thick Ta$_{2}$O$_{5}$ layers were sputtered by reactive high-power impulse magnetron sputtering (HiPIMS) from a Ta target at 6~mTorr pressure, 45~sccm Ar and 5~sccm O$_{2}$ flow rates and 250~W RF power. The thickness and stoichiometric composition of the Ta$_{2}$O$_{5}$ layer were confirmed by XPS spectroscopy. The Ta top electrode and its Pt cap were sputtered on top of the Ta$_{2}$O$_{5}$ film at 4~mTorr pressure, 45~sccm Ar flow and 250~W RF power / 125~W dc power for Ta / Pt, preventing the formation of a native oxide layer at the Ta$_{2}$O$_{5}$/Ta interface. The 2.5~$\mu$m wide bottom and top electrodes were patterned by standard optical lithography and lift-off.

The VO$_{2}$ films were created by the post-deposition heat treatment of a Si/SiO$_{2}$/V structure, where the SiO$_{2}$ / V thickness was 1~$\mu$m / 100~nm. The heat treatment was carried out in air at 400~$^{\circ}$C temperature and 0.1~mbar pressure over 4.5~hours, resulting in a 180~nm thick V$_{2}$O$_{5}$ bottom layer and a 40~nm thick VO$_{2}$ top layer as verified by cross-sectional TEM and EELS analyses \cite{Posa2023}. The metal electrodes consisting of a 10~nm thick Ti adhesion layer and a 50~nm thick Au film were patterned by standard electron-beam lithography and deposited by electron-beam evaporation at 10$^{-7}$~mbar base pressure at rates of 0.1~nm/s and 0.4~nm/s, respectively, followed by lift-off.

\section{Switching dynamics simulation of the Ta$_2$O$_5$ detector circuit}

\begin{figure}[b!]
     \includegraphics[width=0.65\columnwidth]{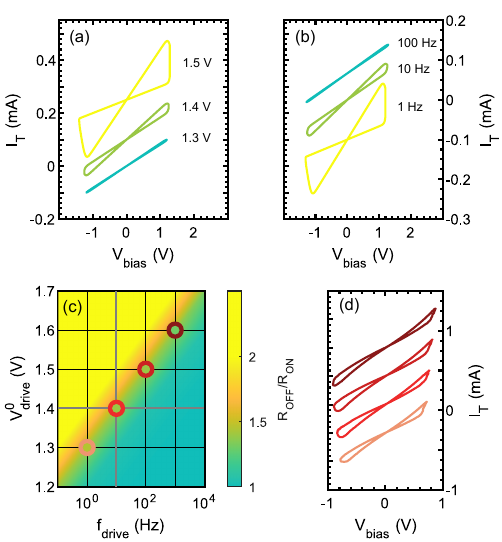}
     \caption{\textbf{The switching dynamics of the Ta$_{2}$O$_{5}$ memristor.} (a) Simulated $I(V)$ traces showing the effect of changing the $V_{\rm drive}^{0}$ amplitude of the triangular driving voltage signal at $f_{\rm drive}$=10~Hz. (b) Simulated $I(V)$ traces illustrating the effect of changing $f_{\rm drive}$ at a constant $V_{\rm drive}^{0}$=1.4~V. (c) Color plot of the $R_{\rm HRS}/R_{\rm LRS}$ ratio deduced from the zero-bias slopes of the simulated $I(V)$ traces as a function of the $V_{\rm drive}^{0}$ amplitude and $f_{\rm drive}$ frequency of the driving voltage signal. The simulations were performed at the discrete frequency and voltage values represented by the grid lines and the axes, while at intermediate values the color scale is interpolated. (d) Measured $I(V)$ traces sharing similar $R_{\rm HRS}/R_{\rm LRS}=1.38\pm0.08$ resistance ratios at various $f_{\rm drive}$ and $V_{\rm drive}^{0}$ settings and $R_{\rm s,T}$=1~k$\Omega$. The driving parameters and the calculated resistance ratios are indicated by the circles in (c) whose outline colors (inner colors) symbolize the colors of the curves (represent their $R_{\rm HRS}/R_{\rm LRS}$ ratios) in (d). Note that the color scale of the simulations in (c) highlights the relevant resistance ratios of the measurements in (d) and saturates at higher resistance ratios. The $I(V)$ curves in (a), (b) and (d) are vertically offset for clarity.}
     \label{fig7}
\end{figure}

Figure~\ref{fig7} shows the results of the simulations in comparison with selected experimental $I(V)$ traces of the Ta$_{2}$O$_{5}$ memristor in series with $R_{\rm s,T}$=1~k$\Omega$. The individual simulated traces in Figs.~\ref{fig7}(a) and (b) illustrate the onset of nonvolatile resistive switching as the $V_{\rm drive}^{0}$ is linearly increased at a constant $f_{\rm drive}$ or, alternatively, as $f_{\rm drive}$ is exponentially decreased at a constant $V_{\rm drive}^{0}$. The color-scale plot in Fig.~\ref{fig7}(c) displays the $R_{\rm HRS}/R_{\rm LRS}$ resistance ratio deduced from the zero-bias slopes of the simulated $I(V)$ traces for the experimentally relevant 1.2~V$<V_{\rm drive}^{0}<$1.7~V voltage interval. Note, that while the latter covers a $\approx$40\% variation the $f_{\rm drive}$ driving frequency spans over 5 orders of magnitude. The colored empty circles in Fig.~\ref{fig7}(c) label the driving parameters of the measured $I(V)$ traces shown in the corresponding colors in Fig.~\ref{fig7}(d). The measured $I(V)$ characteristics in Fig.~\ref{fig7}(d) unambiguously confirm that achieving an identical $R_{\rm HRS}/R_{\rm LRS}$ resistance ratio at linearly increasing voltage requires an exponentially increasing frequency. In the experiment, a constant $R_{\rm HRS}/R_{\rm LRS}\approx$1.5 ratio was chosen, representing the onset of resistive switching. We explored those $V_{\rm drive}^{0}$ and $f_{\rm drive}$ settings which reproduced this ratio. Finally, the simulation parameters were optimized at $A$=0.088~V and $B$=1.06~V by finding the best match between the simulated and measured $I(V)$ traces throughout the investigated $V_{\rm drive}^{0}$\,--\,$f_{\rm drive}$ plane.

\section{The implementation of the neural circuit}

\begin{figure}[b!]
     \includegraphics[width=0.65\columnwidth]{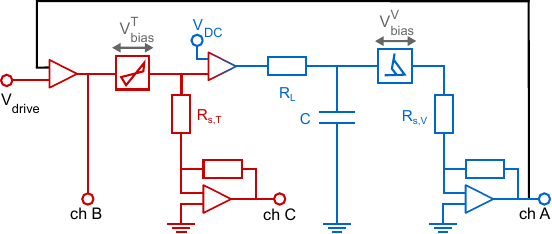}
     \caption{\textbf{The detailed schematic of the artificial neuron circuit.} The $V_{\rm drive}$ input voltage is measured at channel B of a digital storage oscilloscope. The current of the Ta$_{2}$O$_{5}$ / VO$_{2}$ memristor is measured via a current amplifier connected to channel C / channel A of the DSO. The coupling of the detector (red) and oscillator (blue) modules is realized through a voltage amplifier. Through the latter, the additional offset bias $V_{\rm DC}$ is added which sets the operation point of the oscillator module. In order to limit the output pattern of the neuron circuit to a single neural spike and to reset it after a spike detection, the spiking output of the oscillator is fed back to the input of the detector circuit via a second voltage amplifier. The labeled passive circuit components were $R_{\rm s,T}=R_{\rm s,V}$=1~k$\Omega$, $R_{\rm L}$=50~k$\Omega$ and $C$=4.4~nF.}
     \label{fig8}
\end{figure}

The artificial neural circuit consists of two main parts. Its detailed schematic including peripheral instruments is shown in Fig.~\ref{fig8}. The Ta$_2$O$_5$ module (red) is responsible for the detection of the neural spikes in the input signal. The VO$_2$ oscillator module (blue) generates the output spikes. The $V_{\rm drive}$ input voltage pattern of the neuron circuit is provided by adding the off-set spiking output of a Rigol DG5252 arbitrary waveform generator (AWG) and the noise output of a Siglent SDG1050 AWG via a LeCroy DA1850A differential amplifier. $V_{\rm drive}$ is independently recorded at channel B of a Picoscope 6424E digital storage oscilloscope (DSO). The output of the detector module, i.e., the voltage on the $R_{\rm s,T}$ series resistor, is measured via a Femto DLPCA-200 current amplifier (red) at channel C of the DSO. The operation point of the oscillator module is set by the constant voltage output of an Agilent 33220A AWG unit. This constant voltage and the voltage output of the detector module are added and amplified via a Tektronix AM502 differential amplifier (red-blue). Additionally, this unit also incorporates a 100~Hz low-pass filter to remove the noise from the input signal of the oscillator module, greatly improving the reliability of the latter. The output of the oscillator module, i.e., the voltage on the $R_{\rm s,V}$ series resistor, is measured via a second Femto DLPCA-200 current amplifier (blue) at channel A of the DSO. The feedback coupling (black) is incorporated to reset the neural circuit after a detection event, enabling continuous operation. It is implemented through an SRS SR-235 amplifier located at the input of the circuit (red). This unit amplifies the output of the oscillator module and adds it to the input signal of the detector module with a negative polarity. The gain of the feedback must be optimized to match the device characteristics of the Ta$_{2}$O$_{5}$ memristor. A too strong reset pulse increases the risk that the detector circuit will idle in a too high $R_{\rm HRS}$ or exhibit only a smaller, short-lived resistance change upon the arrival of the next neural spike. In contrast, when the gain of the coupling is too low, a single reset pulse may not be sufficient to restore the optimal HRS. Consequently, a single neural spike detection event triggers multiple output spikes.

The circuit with the global feedback resets the detector module autonomously, i.e., this circuit can be fed with a continuous input stream. In our analysis we applied a $50\,$s long input stream of white noise including $50$ Gaussian spikes with $1\,$s repetition rate. Fig. 5(e) demonstrates $10$ different segments of this continuous input stream. The $98\,\%$ detection accuracy was evaluated from the entire stream which was divided to $100$ segments from which 50 (50) segments include (exclude) a Gaussian spike. At the optimized feedback conditions we found that a single output spike of the oscillator circuit was sufficient to reset the Ta$_2$O$_5$ detector module in 46 out of the 50 cases. The low pass filter between the detector and the oscillator module, however, delayed the switch-off of the oscillator circuit yielding multiple output spikes in a portion of the measurements. This delay yielded $1.88$ output spikes on average for the $50$ segments with a Gaussian input spike. 

{\section{{Neural spike detection accuracy at different signal-to-noise ratios}}}

\begin{figure}[h!]
\includegraphics[width=\columnwidth]{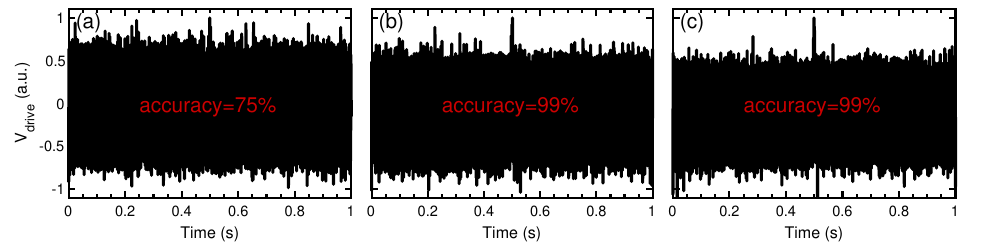}
\caption{{Example time-traces of the driving voltage patterns representing our analysis of the neural spike detection accuracy at different signal-to-noise ratios. Panels (a), (b) and (c) respectively correspond to $V_\mathrm{spike}/V_\mathrm{rmsnoise}=1.3$, $V_\mathrm{spike}/V_\mathrm{rmsnoise}=2$ and $V_\mathrm{spike}/V_\mathrm{rmsnoise}=2.3$.}}
     \label{fig81}
\end{figure}

{We have tested the accuracy of neural spike detection at different signal-to-noise ratios (Fig.~\ref{fig81}). The middle panel (Fig.~\ref{fig81}b) represents the measurement also demonstrated in the main text (Fig.~4f in the main text), where the spike amplitude normalized to the rms noise value is $V_\mathrm{spike}/V_\mathrm{rmsnoise}=2$. Note, that this is already a rather small signal-to-noise ratio, where the amplitude of noise peaks frequently exceeds that of the neural spike. The left panel (Fig.~\ref{fig81}a) demonstrates a significantly smaller signal-to-noise ratio ($V_\mathrm{spike}/V_\mathrm{rmsnoise}=1.3$), where the neural spike is visually hardly distinguishable from the largest noise spikes. In this case the detection accuracy decreases to $75\%$, which is still much better than random guess ($\approx 50\%$  detection accuracy). The right panel (Fig.~\ref{fig81}b) demonstrates a larger signal-to-noise ratio ($V_\mathrm{spike}/V_\mathrm{rmsnoise}=2.3$), where also a high, $99\%$ detection accuracy is obtained. The evaluation protocol is same in the three cases, relying on the analysis on 50 time-traces with neural spikes and 50 time-traces without neural spikes. }

\section{{Time series prediction with two memristors: driving scheme and detailed prediction results}}

\begin{figure}[h!]
\includegraphics[width=\columnwidth]{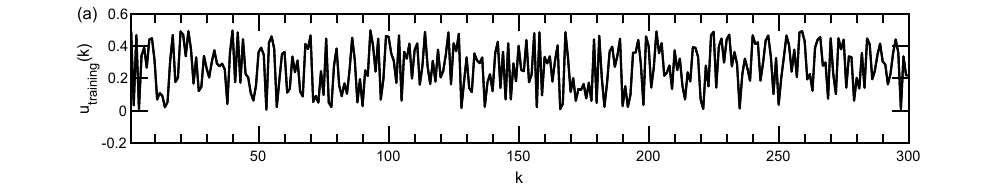}
\includegraphics[width=\columnwidth]{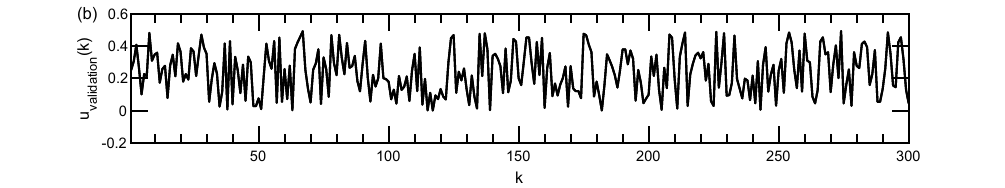}
\caption{{Training (a) and validation (b) datasets used for our time-series prediction analysis.}}
     \label{fig9}
\end{figure}

\begin{figure}[h!]
\includegraphics[width=0.65\columnwidth]{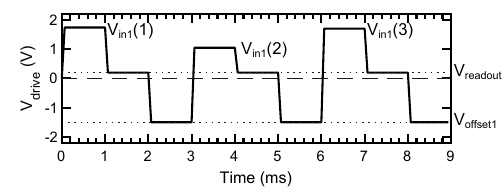}
\caption{{The sequence of the driving signal for our time-series prediction analysis. The input information is encoded in the amplitudes of the positive programming pulses ($V_\mathrm{in1}(k), V_\mathrm{in2}(k))$, the system response is read out along the readout intervals ($V_\mathrm{readout}$), and the forgetting is achieved by the tunable negative offset voltages ($V_\mathrm{offset1}, V_\mathrm{offset2}$). The figure illustrates the first three periods of the first channel's driving sequence.}}
\label{fig10}
\end{figure}

{For the time-series prediction task we have generated the $u_\mathrm{training}(k)$ (Fig.~\ref{fig9}a) and $u_\mathrm{validation}(k)$ (Fig.~\ref{fig9}b) training and validation input data sets, both including 300 uniformly distributed random numbers. From these numbers the training and validation waveforms were generated with a Rigol DG5252 arbitrary waveform generator following the scheme in Fig.~\ref{fig10}. The $V_\mathrm{in1}(k)=V_\mathrm{in2}(k)=u(k)\cdot a + b$ input voltages (see Fig.~6a in the main text) are encoded in the amplitudes of the positive voltage pulses in the 1$^\mathrm{st}$ $1\,$ms long segment of the $3\,$ms long driving period, afterwards a $1\,$ms long readout voltage is applied ($V_\mathrm{readout}=200\,$mV), and finally a negative $V_\mathrm{offset1}$ or $V_\mathrm{offset2}$ offset voltage is applied to induce forgetting. Figure \ref{fig10} illustrates the first three periods of the driving sequence applied on the first input channel of the memristive circuit. Both input channels are driven by this scheme applying different $V_\mathrm{offset1}$ and $V_\mathrm{offset2}$ offset voltages.  The $V_\mathrm{out1}(k)$ and $V_\mathrm{out2}(k)$ output voltages are evaluated as the average output voltage measured along the given readout voltage segment.}

\begin{figure}[h!]
\includegraphics[width=\columnwidth]{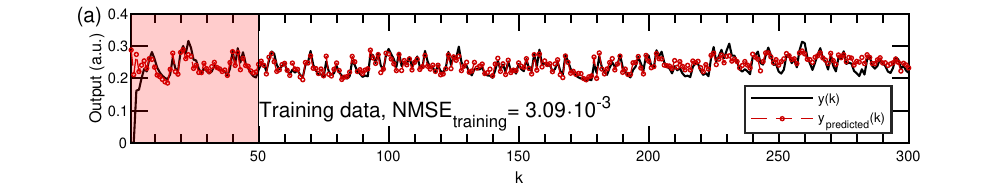}
\includegraphics[width=\columnwidth]{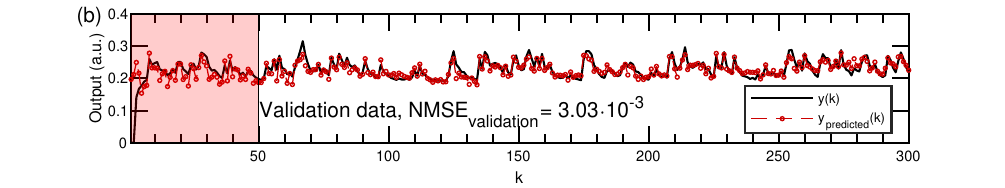}
\caption{{Output streams as predicted by the memristive dynamical circuit with two memristors (red circles) in response to the training (a) and validation (b) datasets. The predicted streams are compared to the true $y(k)$ outputs of the mathematical dynamical system (black lines).}}
     \label{fig11}
\end{figure}

{In our analysis we measure the response of both memristive channels to both the training and the validation waveforms. The response to the training waveform is used to find the optimized offset values. At a certain $V_\mathrm{offset1}, V_\mathrm{offset2}$  setting the most predictive linear combination of the two physical output channels (i.e. the optimal $w_1, w_2$ and $o$ values) are determined in software using the Adam optimization algorithm, similarly to {Ref.~\citenum{Du2017}} . Finally, the best performing $V_\mathrm{offset1}, V_\mathrm{offset2}$ pair is chosen. Fig.~\ref{fig11}a demonstrates the predicted output stream ($y_\mathrm{predicted}(k)$, red circles) in response to the training waveform using the optimized offset value pair. This is compared to the true $y(k)$ output of the mathematical dynamical system (black line). Following the protocol in {Ref.~\citenum{Du2017}}, we drop the first 50 points from the analysis (red region in the figure), where the $y(k)$ true output of the mathematical dynamical system exhibits an initial transient. This means, that the normalized means squared error is calculated for the last 250 points of the output datastream. For the measurement demonstrated in Fig.~\ref{fig11}a this yields an NMSE$_\mathrm{training}=3.09\cdot10^{-3}$ normalized prediction error. Fig.~\ref{fig11}b represents the predicted output of the same two memristors using the settings optimized along the training process, but measuring the prediction error in response to the independent validation dataset. This figure shows the total measurement from the same data,
of which a shorter, more visible section is shown in Fig.~6b in the main text. Again, the last 250 points are used to evaluate the error, yielding NMSE$_\mathrm{validation}=3.03\cdot10^{-3}$.}

{To explore the possibility for further improvement of the prediction error, a different approach was also applied. The time-series prediction operation was tested spanning an even broader parameter set, such that the programming dynamics was also independently adjustable for the two channels, i.e., different $V_\mathrm{in1}(k)=a_1\cdot u(k) +b_1$ and $V_\mathrm{in2}(k)=a_2\cdot u(k) +b_2$ scaling parameters were applied. Meanwhile, the $V_\mathrm{offset1},V_\mathrm{offset2}$ parameters (i.e. the forgetting dynamics) were scanned with an even finer resolution around the optimal parameter values. In this case, however, solely the response of a single memristor was tested at $100$ different parameters sets. From these response curves we have assembled the combined response of the two memristive channels in software, testing all the $10^4$ possible parameter pairs. With this approach a further improved performance was achieved yielding NMSE$_\mathrm{training}=2.79\cdot10^{-3}$ and NMSE$_\mathrm{validation}=2.47\cdot10^{-3}$.}

\begin{figure}[h!]
\includegraphics[width=\columnwidth]{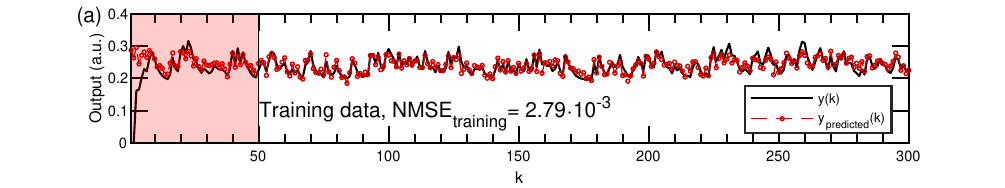}
\includegraphics[width=\columnwidth]{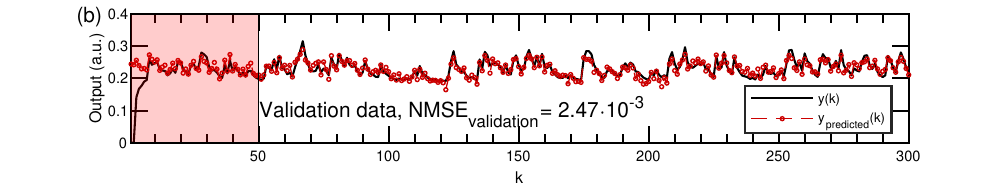}
\caption{{Further optimized system performance by a broader and more refined scanning of the parameter space also allowing independently adjustable $a_1, b_1$ and $a_2, b_2$ scaling factors between $u(k)$ and the $V_\mathrm{in1}, V_\mathrm{in2}$ physical voltage inputs of the two channels. In this case, however, the response at 100 different parameter settings was measured on a single memristor, and the response of the two channels at a certain parameter pair was calculated in software. Panels (a) and (b) demonstrate the predicted outputs (red circles) in comparison to the true $y(k)$ output (black lines) for the training and the validation data, respectively. The pulse amplitudes of the first and second memristive channels were calculated as $V_{\rm in1}=(u\cdot 1.5+1.8)$~V and $V_{\rm in2}=(u\cdot 2+0.9$)~V. The optimized negative offset values were $V_{\rm offset1}=-2.3$~V and $V_{\rm offset2}=-1.8$~V, respectively. In this case the driving protocol was somewhat modified lacking the readout voltage segment, and reading the device state during the last $0.1\,$ms portion of the programming pulses. Accordingly, the output stream was calculated by normalizing the outputs to the input levels, $y_\mathrm{predicted}(k)=(V_\mathrm{out1}(k)/V_\mathrm{in1}(k))\cdot w_1+(V_\mathrm{out2}(k)/V_\mathrm{in1}(k))\cdot w_2+o$.}}
     \label{fig12}
\end{figure}

\bibliographystyle{achemso}
\bibliography{ReferencesSupp}